\documentstyle[preprint,aps,prd]{revtex}

\begin{document}

\title {Influence functional in two dimensional dilaton gravity}    

\author{Fernando C.\ Lombardo \footnote{Electronic address: 
lombardo@df.uba.ar} and Francisco D.\ Mazzitelli 
\footnote{Electronic address: fmazzi@df.uba.ar}}

\address{{\it
Departamento de F\'\i sica, Facultad de Ciencias Exactas y Naturales\\ 
Universidad de Buenos Aires - Ciudad Universitaria, 
Pabell\' on I\\ 
1428 Buenos Aires, Argentina}}

\maketitle

\begin{abstract}
We evaluate the influence functional for two dimensional
models of dilaton gravity. This functional is exactly computed when 
the conformal invariance is preserved, and it can be written as the 
difference between the Liouville actions on each 
closed-time-path branch plus a boundary term. From the influence 
action we derive the covariant form 
of the semiclassical field equations. We also study the quantum to 
classical transition in cosmological backgrounds. In the conformal case
we show that the semiclassical approximation is not valid because
there is no imaginary part in the influence action. Finally we show
that the inclusion of the dilaton loop in the influence functional breaks 
conformal invariance and ensures the validity of the semiclassical 
approximation.
\end{abstract}
\newpage
\section{Introduction}

In semiclassical and quantum gravity it is of interest to compute 
the backreaction of quantum fields on the spacetime geometry. One of
the most interesting questions to answer 
in this context is about the endpoint of
black hole evaporation and the information loss puzzle. It would
also be useful to understand which is the effect of the quantum 
fluctuations on classical singularities of general relativity, in 
particular in cosmological settings.

The standard approach to this problem consists of studying
the so called semiclassical Einstein equations, which include 
the quantum effects by taking as a source the quantum mean value
of the energy momentum tensor of the quantum fields. 
The spacetime metric is considered a classical object.
The analysis of
the semiclassical equations in realistic models is a very difficult task. 
For this reason, it is
of interest to analyze solvable toy models in which some of the 
difficulties are not present. Two dimensional dilaton gravity
theories are very useful in this sense. From them one may take a better 
understanding of the main 
aspects about the quantum properties of black holes and the influence of 
quantum effects in cosmological situations \cite{review}.

The fundamental result of
Callan et al \cite{cghs} is that two dimensional gravity coupled to a 
dilaton field $\phi$ and $N$ conformal fields $f_i$ contains black
hole solutions and Hawking radiation. The backreaction of the quantum 
fields
$f_i$  can also be computed in this model. In particular, with
minor modifications \cite{rst,bpp} the semiclassical problem can be 
completely solved.
But {\it is the semiclassical approximation justified}? 
This is of course an
old and important question for which, as we will show, two dimensional
models also provide interesting 
simplifications.

After Hartle and Hawking proposal for the
wave function of the Universe, the validity of the semiclassical
approximation  has been extensively studied,
mainly in four dimensional, cosmological minisuperspace models. It
was realized
that the semiclassical limit is based on two main ingredients: correlations 
and decoherence \cite{jppsinha}. The correlations between different 
variables was first analyzed using the Wigner function \cite{hallicorr}, 
while the decoherence was studied through the reduced density matrix 
\cite{hallideco}. Both ingredients are not independent: and excess of 
decoherence can destroy the correlations \cite{laflame}.
The quantum to classical transition was subsequently analyzed
using the decoherence functional of Hartle and Gell-Mann 
\cite{hartlegellmann},
which is a functional of two histories ${\cal D}[g_{\mu\nu}^+,
g_{\mu\nu}^-]$ after integration of 
the quantum variables. The metric $g_{\mu\nu}$ can be described
as a classical one if the decoherence functional is approximately
diagonal. When this is the case, probabilities for different histories
satisfy the sum rules, and then quantum effects are negligible.

A better understanding of this problem can be achieved by viewing
it in the context of quantum open systems, where the metric
$g_{\mu\nu}$ is viewed as the ``system" and the quantum fields 
as the ``environment". 
The influence functional \cite{feynmanvernon} technique is the adequate 
tool to describe the effect of the environment,
and provides information about dissipation, noise and the quantum
to classical transition suffered by the system. The influence functional
is closely related to the decoherence functional of Hartle and
Gell-Mann, and also gives the temporal evolution of the reduced density
matrix. In summary, is 
a fundamental tool to develop a systematic
study of the validity of the semiclassical approximation.  

In this paper, as a first step towards the analysis of the quantum to 
classical transition in two dimensional models,  
we will compute the influence functional by tracing 
out the quantum matter fields. We will
show that these toy models are very useful to understand
its main features since, due to conformal invariance, it can be
computed exactly. We will use the general result for the influence 
functional to derive the semiclassical equations of the model 
and to discuss the validity of the semiclassical limit in the 
cosmological context. 
We will show that the semiclassical approximation is not 
justified in this case, unless conformal invariance 
is broken, and that this can be done by including 
the quantum fluctuations of the dilaton field (in this case the 
influence functional can be computed only perturbatively). We will
also discuss the dependence of the influence functional
with the matching hypersurface.

The paper is organized as follows. In the next section we introduce
the model, the definition of the influence 
functional and its relation with the Close Time Path (CTP)
effective action \cite{ctp}. In Section III we derive the 
influence functional from the Euclidean effective action. 
In Sec. IV, we derive the semiclassical, covariant 
equations of motion from the CTP effective 
action. Section V contains
the complete evaluation of the influence 
functional for cosmological metrics.
We will find  that for $t=const$ matching hypersurfaces, there is no 
loss of coherence. A discusion about 
an alternative choice of matching surfaces completes this section.
In Sec. VI, 
we will show that the inclusion of the quantum fluctuations of the 
dilaton field breaks conformal invariance, and produces an 
explicit imaginary part in the influence functional that 
reveals the presence of 
particle creation and decoherence effects. Section VII contains our 
final remarks.

\section{The model}

The classical 2D Callan-Giddings-Harvey-Strominger 
(CGHS) action is given by
\begin{equation} S_{CGHS}= {1\over{2 \pi}} \int d^2x \sqrt{-g(x)} 
\{e^{-2\phi}[R + 4 (\nabla \phi)^2 + 4 \lambda^2] - {1\over{2}} 
\sum_{i = 1}^{N} (\nabla f_i)^2\},\label{cghs}\end{equation}
where $\phi$ is the dilaton field, $R$ is the 2D Ricci scalar, 
$\lambda$ is a positive constant, and the $f_i$ are N 
massless scalar matter fields conformally coupled to the 2D geometry.  

Considering the quantum effects of the scalar fields, the exact 
{\it Euclidean} effective action for the two dimensional 
model is \cite{cghs}
\begin{equation} S_{eff}^E= S_{CGHS}^E - {N\over{96 \pi}}\int 
d^2x \sqrt{g(x)}\int d^2x' \sqrt{g(x')} R(x) {1\over{\Box}} 
R(x'). \label{effact}\end{equation}  
The second term in Eq. (\ref{effact}) is the Polyakov-Liouville effective 
action \cite{polia} derived from the trace anomaly of the massless 
scalar fields. This term is non local and contains the 
inverse of the d'lambertian, i.e. the two-point 
Euclidean propagator. In the conformal gauge this effective action can 
be written as

\begin{equation}S_{eff}^E=S_{CGHS}^E + S_{Liouv}^E = S_{CGHS}^E -
{N\over{12}}
\int d^2x
~ \rho ~\partial_+\partial_- ~ \rho .\end{equation}

It is usually claimed that the semiclassical
field equations can be derived from the effective action (\ref{effact}). 
Strictly speaking, this is not correct. 
Replacing the Euclidean propagator by the Feynman one, one
obtains the usual {\it in-out} effective action. As is very
well known, the effective equations derived from this action are neither 
real
nor causal because they are equations for {\it in-out} matrix
elements and not for mean values.
The solution to this problem is also well known. Using the CTP
 formalism \cite{calzhu94}
one can construct an {\it in-in} effective action that produces
real and causal field equations for {\it in-in} expectation values. The 
effective action can be written as (we are assuming a semiclassical
point of view because we are not integrating over the metric 
configurations)
\begin{equation} e^{i S_{eff}^{CTP}[g^{\pm},f^{\pm}]} = {\cal N} 
e^{i(S_{CGHS}[g^+
, f^+]
-S_{CGHS}[g^-,f^-])} 
\int {\cal D} {\hat f}^+{\cal D} {\hat f}^- e^{i(S_{matter}[g^+,{\hat f}^+]
- S_{matter}[g^-,{\hat f}^-])}. 
\label{ctpeff}\end{equation} 
The field equations are obtained taking the 
variation of this action with respect to the $g_{\mu\nu}^+$ metric, and 
then 
setting $g_{\mu\nu}^+=g_{\mu\nu}^-$. The integration in Eq. (\ref{ctpeff}) 
is over quantum 
fluctuations around the background matter fields. ${\hat f}^+$ and 
${\hat f}^-$ must contain negative 
and positive frequency modes, respectively, in the remote past 
(these are the {\it in} boundary conditions) and must coincide at a 
finite future spacelike hypersurface. This hypersurface, that must be 
a Cauchy hypersurface, will be denoted by $\Sigma$. The 
path integral can be thought as the 
path sum of two differents fields evolving in two temporal branches; 
one going forward in time in the presence of source $g_{\mu\nu}^+$ 
from the 
{\it in} vacuum to $\Sigma$ and the other 
backward in time in the presence of $g_{\mu\nu}^-$ from $\Sigma$ to the 
{\it in} vacuum. 
The constraint that must be imposed is ${\hat f}^+\vert_{\Sigma}={\hat f}^
-\vert_{\Sigma}$. We stress that
${\hat f}$ and $g_{\mu\nu}$ 
are independent on the + and $-$ branches. 

It can be easily proved that
the CTP-effective action takes the form
\begin{equation}S_{eff}^{CTP}= S_{CGHS}[g_{\mu\nu}^+,f^+]-S_{CGHS}
[g_{\mu\nu}^-,f^-]
+\Gamma_{IF}[g_{\mu\nu}^{\pm}].\label{if}\end{equation} The functional 
$\Gamma_{IF}[g_{\mu\nu}^{\pm}]$ is so called {\it 
influence action} \cite{calzhu94}. Note that it depends (non trivially,
as we will see) on the matching hypersurface $\Sigma$. From the 
quantum open systems point of view, after integration of the 
``environment" (the quantum fluctuations of the matter fields ${\hat f}_i$), 
one 
ends up with an 
effective theory for the ``system" (the metric $g_{\mu\nu}$, the dilaton, 
and the classical background of the matter fields, $f_i$). The 
quantity  $e^{iS_{eff}^{CTP}}$ is the influence 
functional and coincides with the
{\it decoherence functional} of Hartle and Gell-Mann \cite{hartlegellmann}. 

In our 
present case, we are choosing an initial condition such that 
the initial quantum state for the scalar fields is the {\it in}  vacuum
state. With this 
particular choice there is a simple relation between the influence functional 
and the effective action, as can be seen 
from Eq. (\ref{if}). In general, the influence functional is a more 
complicated object that strongly depends of the initial 
conditions \cite{cgqft}.

It is interesting to note that the influence functional
can be written as 
\begin{equation} e^{iS_{eff}^{CTP}}=\sum_{\alpha}\langle 0, in\vert\alpha,T
\rangle_{g^-}\langle \alpha ,T\vert 0, in\rangle _{g^+},\label{fieldprod}
\end{equation} 
therefore, it can be interpreted as the scalar product 
on $\Sigma$ between the states constructed 
as temporal evolutions (on the two different metrics
$g_{\mu\nu}^{\pm}$) from the common $in$ state up to the future 
hypersurface $\Sigma$.
The CTP-effective action 
can alternatively written in terms of the Bogolubov coefficients 
connecting the $in$ and $out$ basis
in each temporal branch.
This implies that there is 
decoherence if and only if there is particle creation during the 
field evolution \cite{calzmazzi}.

\section{Two dimensional Influence Functional}

In an alternative, and more concise notation, we can write the effective
action of Eq. (\ref{ctpeff}) as \cite{mottola,nos}
\begin{equation} e^{i S_{eff}^{{\cal C}}[g,f]} = {\cal N} 
e^{i S_{CGHS}^{{\cal C}}[g, f]} 
\int {\cal D} {\hat f} e^{i S_{matter}^{{\cal 
C}}[g,{\hat f}]},\label{neweff}\end{equation} where we have introduced the 
CTP complex temporal path ${\cal C}= {\cal C}_+ \cup {\cal C}_-$, going 
from minus infinity to $\Sigma$ (${\cal C}_+$), 
and backwards, with a decreasing (infinitesimal) imaginary part 
($\cal C_-$). 
Time integration over the contour ${\cal C}$ is defined by $\int_{{\cal C}} 
dt =\int _{{\cal C_+}} dt -\int_{{\cal C_-}} dt$. The field fluctuation 
${\hat f}$  appearing in 
Eq. (\ref{neweff}) is related to those in Eq. (\ref{ctpeff}) by 
${\hat f}(t, x) = 
{\hat f}_{\pm}(t,x)$ if $t \in {\cal C}_{\pm}$. The same applies to 
$g_{\mu\nu}$ and to the background $f$. This equation is 
useful because it has the 
structure of the usual {\it in-out} or the Euclidean effective action. 
Feynman
rules are therefore the ordinary ones, replacing Euclidean propagator by
\begin{eqnarray} G(x,y) = \left\{\begin{array}{ll} 
G_{++}(x,y)=i \langle 0, in\vert T {\hat f}^+(x) {\hat f}^+(y)\vert 0, 
in\rangle,& ~t, t' ~ 
\mbox{both on} ~{\cal C}_+ \\ G_{--}(x,y)=-i \langle 0, in\vert {\tilde T}
{\hat f}^-(x) 
{\hat f}^-(y)\vert 0, in\rangle , & ~t, t' ~ \mbox{both on} 
~{\cal C}_-\\ G_{+-}
(x,y)=-
i \langle 0, in\vert {\hat f}^+(x) {\hat f}^-(y)\vert 0, in\rangle,    
&~t ~\mbox{on}~ 
{\cal C}_+, t'  ~\mbox{on} ~{\cal C}_-\\ G_{-+}(x,y)=i \langle 0, in
\vert {\hat f}^-(y) 
{\hat f}^+(x)\vert 0, in\rangle, & ~ t  ~\mbox{on} ~ {\cal C}_-, t'~  
\mbox{on}~ 
{\cal C}_+\end{array}\right. \label{prop} \end{eqnarray} 

It is important to note that each propagator has been defined taking 
into account that the field fluctuations correspond to different 
metrics. For example, $G_{++}$ is the time ordered 
product for  both fields in the + metric, while $G_{--}$ is the anti-temporal 
ordered product for fields in the $-$ metric. Therefore, in this case it 
is not true 
that $G_{++} = G^*_{--}$ because each propagator is defined on a different 
metric. In a similar way, $G_{+-}$ contains one field on the + metric 
and the other one on the $-$ metric. Moreover, the Feynman and Dyson 
propagators can be expresed as
\begin{equation}G_{\pm\pm}(x,y)=G_{\pm}(x,y)~\theta(x^0-y^0) + G_{\mp}(x,y)
~\theta(y^0-x^0)\end{equation}
but now it is not true that $G_{+}=G_{+-}$ and $ G_{-}=G_{-+}$. 

From Eqs (\ref{effact}) and (\ref{prop}) the closed-time-path effective 
action can be written as
\begin{eqnarray}S_{eff}^{CTP} &=& S_{CGHS}(g_{\mu\nu}^+, f^+) - S_{CGHS}
(g_{\mu\nu}^-, f^-)\nonumber \\ &-& {N\over{96 \pi}}\int d^2x\sqrt{-g(x)}
\int d^2y\sqrt{-g(y)}~~R^a(x)~~ G_{ab}(x,y)~~R^b(y),\label{explctp}
\end{eqnarray}
where the indices $a$ and $b$ are denoting each CTP branch + and $-$. From 
this expression is possible, at least formally, to write the real 
and imaginary parts 
of the influence action for two generic +/$-$ metrics.

It is instructive to analize the influence functional in the 
conformal gauge        
\begin{equation}g_{+-}= -{1\over{2}}e^{2 \rho} ~~~~, ~~~~ g_{--}=g_
{++}=0.\end{equation}
The Ricci scalar is $R= - 8 e^{-2 \rho}\partial_+\partial_-\rho$,
where $\partial_{\pm}$ denotes derivatives with respect to the
coordinates   $x^{\pm}=
(x^0 \pm x^1)$.
In this gauge, the closed-time-path-effective action 
reads
\begin{eqnarray}S_{eff}^{CTP}[\rho^+,f^+,\rho^-,f^-;\Sigma ]&=& 
S_{CGHS}(\rho^+,f^+) - S_{CGHS}(\rho^-,f^-)\nonumber \\
&-&{N\over{6 \pi}}\int d^2x\int d^2y ~~\partial_+\partial_-\rho^
a(x^{\pm}) ~~G_{ab}(x^{\pm},y^{\pm})~~\partial_+\partial_-\rho^b(y^{\pm})
.\label{cgctpefa}\end{eqnarray}

After integrations by parts, the 
effective action can be expressed as the classical 
terms, plus the 
difference between the Liouville action in each metric, plus boundary 
terms
\begin{eqnarray}S_{eff}^{CTP}[\rho^+,f^+,\rho^-,f^-;\Sigma ]&=& 
S_{CGHS}(\rho^+,f^+) - S_{CGHS}(\rho^-,f^-)\nonumber \\
&-&{N\over{12 \pi}}\int d^2x  ~~ [\rho^+\partial_+\partial_-\rho^+  
- \rho_- \partial_+\partial_-\rho^-]\nonumber \\
&-&{N\over{6 \pi}}\{\mbox{boundary terms}\},\label{supterms}\end{eqnarray}
where
\begin{eqnarray}\{\mbox{boundary terms}\}&=&  \int_{-\infty}^{+\infty} 
dx \int_{-\infty}^{+\infty} dy \left[\partial_{x^-}\Delta (x,k(x))~ 
N_1[x,k(x);y,{\bar k}(y)]
~ \partial_{y^-}\Delta (y,{\bar k}(y))\right. \nonumber \\
&+& \left. 2 \partial_{x^-}\Xi (x,k(x))~ N_2[x,k(x);y,{\bar k}(y)] ~ 
\partial_{y^-}\Delta (y,{\bar k}(y)) 
\right. \nonumber \\
&+& \left. \partial_{x^-}\Xi (x,k(x))~ N_3[x,k(x);y,{\bar k}(y)]~ 
\partial_{y^-}\Xi (y,{\bar k}(y))\right]
\nonumber \\
&-& 2\int_{-\infty}^{+\infty} dx \int_{-\infty}^{+\infty}dy 
\left[\Xi (x,k(x)) 
~ \partial_{x^+} N_2[x,k(x);y,{\bar k}(y)] ~ \partial_{y^-} 
\Delta (y,{\bar k}(y))\right. 
\nonumber \\
&+&  \left.\Xi (x,k(x)) ~ \partial_{x^+} N_4[x,k(x);y,{\bar k}(y)] ~ 
\partial_{x^-} \Xi (y,{\bar k}(y))
\right. \nonumber \\
&-& \left. {1\over{2}}\Xi (x,k(x))~ \partial_{x^+} \partial_{y^+} 
N_4[x,k(x)
;y,{\bar k}(y)]
~ \Xi (y,{\bar k}(y))\right]\label{exbt}.\end{eqnarray}
The matching 
hypersurface $\Sigma$ is defined by $t_x=k(x)$, $t_y={\bar k}(y)$; 
$\Delta ={1\over{2}}( \rho^+-\rho^-)$, $\Xi = {1\over{2}}(\rho^+ 
+ \rho^-)$, and 
\begin{eqnarray}N_1&=&G_{++}+G_{+-}-G_{-+}-G_{--}\nonumber \\
N_2&=&G_{++}+G_{+-}+G_{-+}+G_{--}\nonumber \\
N_3&=&G_{++}-G_{+-}-G_{-+}+G_{--}\nonumber \\
N_4&=&G_{++}-G_{+-}+G_{-+}-G_{--}.\end{eqnarray}

The expression (\ref{supterms}) for the effective action is  
absolutely general, and can be applied to any particular metric in the 
conformal gauge. 
It is worth to note that if the influence action contains
a non-trivial imaginary contribution, it must be included in the boundary
terms.
If both metrics $g_{\mu\nu}^{\pm}$ coincide asymptotically in the 
future, and if the matching
hypersurface is within such region, all the boundary terms 
vanish because the usual relations 
between Green functions are valid as in the remote past; $N_4$ and 
$\Delta$ are simultaneously zero and only the trace 
anomaly survives. 

The main result of this Section is that the influence action can be
easily computed from the Euclidean effective action. The matching
hypersurface $\Sigma$ plays a crucial role in its evaluation. In 
particular, in the conformal gauge all the relevant information about
the quantum to classical transition is contained in a $\Sigma$-dependent 
boundary term.

\section{The covariant equations of motion}

Although we are mainly interested in the analysis of the 
quantum to classical transition in the two dimensional models, 
in this Section we will derive the covariant field equations
from the CTP effective action. In previous works \cite{example},
it was claimed that the the semiclassical equations of motion
follow from the usual effective action Eq. (\ref{effact}). As we 
already 
pointed out, strictly speaking this is not correct, since there is no 
variational
principle for the initial-value problem, unless one
uses the CTP formalism. Indeed, the semiclassical 
field equations have been previously obtained  from the classical
equations by  taking as a source the 
quantum mean value of the energy-momentum tensor, i.e.,
\begin{equation}
2 \pi {\delta S_{CGHS}
\over{\delta g_{\mu\nu}}} = \langle T_{\mu\nu}\rangle.
\label{scghse}
\end{equation}
The components of the  energy-momentum tensor have been derived
from the trace anomaly and the imposition of the conservation laws
$\langle T^{\mu\nu}\rangle_{;\nu} = 0$ \cite{christensen}. In all 
previous
works the field equations have been written and analyzed 
in the conformal gauge. Moreover, with this procedure it is
possible to get the field equations only for conformal matter
fields, since for massive or non conformally coupled 
fields the trace of the energy momentum tensor is not known
a priori.

The CTP formalism allow us to derive the covariant 
equations of motion from
\begin{equation}{\delta S_{eff}^{CTP}\over{\delta g_{\mu\nu}^+}}
\vert_{g_{\mu\nu}^+
=g_{\mu\nu}^-}
= 0.\label{fieldequation}\end{equation}
At this point, the only difficulty resides in the knowledge 
of the functional variation 
of the Green functions with respect to the metric. From the definition 
of the Green functions, and after expanding the field in modes we can
prove that \cite{dewitt}
\begin{equation}\delta G_{++}= G_{ret}~~ \delta \Box ~~ G_{++}+ G_{++}~~ 
\delta \Box~~ G_{adv}- G_{ret}~~ \delta \Box ~~ G_{adv},\label{g++}
\end{equation} 
\begin{equation}\delta G_{+-}= G_{ret}~~ \delta \Box ~~ G_{+-},\label{g+-}
\end{equation}
\begin{equation}\delta G_{-+}= G_{ret}~~ \delta \Box ~~ G_{-+},\label{g-+}
\end{equation}
where $G_{ret}$ and $G_{adv}$ are the usual retarded and advanced 
Green functions, and where 

\begin{equation}\delta \Box = - \nabla^\mu\nabla^\nu \delta g_{\mu\nu}
-{1\over{2}} \partial^\lambda g^{\mu\nu}\left( \delta g_{\lambda\nu;\mu}
+\delta g_{\mu\lambda;\nu} - \delta g_{\mu\nu;\lambda}\right).\label{dalvar}
\end{equation} 
In Eqs. (\ref{g++}) - (\ref{dalvar}) all the propagators $G_{ret}$, $G_{adv}$
, $G_{+-}$, and $G_{-+}$ are evaluated at $g_{\mu\nu}^+ = g_{\mu\nu}^-$, since
  this is all we need to obtain the field equations (see Eq. 
(\ref{fieldequation})). 

After some algebra, the covariant equations of 
motion can be written as
\begin{eqnarray}{\delta S_{CGHS}\over{\delta g_{\mu\nu}^+}} \equiv
&& {1\over{2\pi}}\langle T_{\mu\nu} \rangle = \nonumber \\
&-&{N\over{48 \pi}}\int d^2y\sqrt{-g(y)} ~R(y)~[\nabla_{\mu} 
\nabla_{\nu} - g_{\mu\nu}\Box ]_{(x)} ~G_{ret}(x,y)]\nonumber \\
&+&{N\over{192 \pi}}\int d^2x\sqrt{-g(x)} \int d^2y\sqrt{-g(y)}\left\{
2 R(x)~G_{ret}(x,z)\right. \nonumber \\
&&\times ~ [\nabla_{\mu} \nabla_{\nu} - {1\over{2}}g_{\mu\nu}
(z)\Box ]_{(z)} ~G_{ret}(z,y)~R(y)\nonumber \\
&-&~2 R(x) ~ \partial_\mu (z)G_{ret}(x,z)~ \partial_\nu (z)G_{ret}(z,y)
~ R(y)\nonumber \\
&+&\left. R(x) ~ g_{\mu\nu}(z) ~ \partial^{\alpha}(z) G_{ret}(x,z)
~ \partial_{\alpha}(z) G_{ret}(z,y)~R(y)\right\}  .
\label{eqmotion}\end{eqnarray}
Note that these are non-local, real, and causal equations of motion. From 
these equations we can calculate the trace of the stress 
tensor
\begin{equation}\langle T_{\mu}^{\mu}\rangle = 2\pi g^{\mu\nu}{\delta 
S^{CTP}_{eff}\over{\delta g_{\mu\nu}^+}}= N ~ {R\over{24}},\label{trace}
\end{equation}
which gives the well known trace anomaly \cite{birrel&davies}. 

To make contact with previous works, we write the covariant equations 
of motion in the conformal 
gauge. Writting the curvature scalar in this gauge, the 
components of the stress 
tensor read
\begin{equation}\langle T_{+-}\rangle=\langle T_{-+}\rangle = -
{N\over{12}}\partial_+\partial_-\rho,\label{tmasmenos}\end{equation}
\begin{equation}\langle T_{\pm\pm}\rangle = -{N\over{12}}[\partial_\pm
^2\rho - \partial_\pm \rho \partial_\pm \rho - t^\pm].\label{tmasmas}
\end{equation}
The functions $t^{\pm}$ depend on $x^\pm$ and can be 
expressed as
\begin{equation}t^{\pm}= \partial_{\pm}^2 S^{\pm} - 2 \partial_{\pm}
S^{\pm}\partial_{\pm}S^{\pm},\end{equation} 
where functions $S^{\pm}$ are given by
\begin{equation}S^{+}(x^+) = \int d^2y ~\partial_{y^-} [\rho (y) 
\partial_{y^+} G_{ret}(x,y)],\end{equation} 
\begin{equation}S^-(x^-) = \int d^2y ~\partial_{y^+} [\partial_{y^-} 
\rho (y) G_{ret}(x,y)].\end{equation}
Of course, Eqs. (\ref{tmasmenos}) and (\ref{tmasmas}) coincide with the 
results obtained in previous works \cite{cghs,rst} using a different method.

The functions $t^{\pm}$ depend on the quantum state of the matter fields. 
In our case, they are completely determined by the  boundary
conditions in the remote past that we used to compute the CTP
effective action (see Eq. (\ref{ctpeff})), and correspond to the
$in$ vacuum state. 
As an example, we will obtain the explicit expression of 
these functions for cosmological metrics, and
for a collapsing matter wave. 

For cosmological metrics, $\rho$ is a function of the conformal time $t$, 
$\rho = \rho (t)$.
Then using that the retarded propagator is given by 
\begin{equation}G_{ret}(x,y)= \theta (t_x - t_y - \vert x 
- y\vert ),\end{equation}
we can compute the functions $S^{\pm}$ and $t^\pm$, that in this case are
\begin{equation}t^{\pm}=\partial_\pm^2 \rho - \partial_\pm \rho 
\partial_\pm \rho.\end{equation}
Therefore, from Eqs. (\ref{tmasmenos}) and (\ref{tmasmas}) we  
obtain 
\begin{equation}\langle T_{+-}\rangle=\langle T_{-+}\rangle = -
{N\over{12}}\ddot{\rho},\end{equation}
\begin{equation}\langle T_{\pm\pm}\rangle = ~0.\end{equation}
Only the trace anomaly survives \cite{russomazzi}.

Let us now consider the well known case of an $f$ shock wave 
traveling in 
the $x^-=t - x$ direction described by the stress tensor
\begin{equation}{1\over{2}}\partial_+f \partial_+f = a \delta (x^+ - x^+_0)
,\end{equation} 
where $a$ is a positive constant. The classical metric is
given by 
\begin{equation}e^{-2\rho}=e^{-2\phi}=-a (x^+-x^+_0)\theta(x^+ -x^+_0)
- \lambda^2 x^+ x^-.\end{equation} 
For $x^+ < x^+_0$, this is simply 
the linear dilaton vacuum, while for $x^+ > x^+_0$ there is a black hole 
with mass $a x^+_0 \lambda$. 

If we introduce the ``tortoise'' coordinates $\lambda \sigma^{\pm}$
\begin{equation}\lambda x^+=e^{\lambda\sigma^+} ~~~ , ~~~ \lambda x^- 
= -(e^{-\lambda \sigma^-}+{a\over{\lambda}}),\end{equation} 
the metric can be written as 
\begin{eqnarray} e^{-2\rho}= \left\{\begin{array}{ll} 
[1 + ({a\over{\lambda}})e^{\lambda \sigma^-}]^{-1},& ~ 
\mbox{if} ~\sigma^+ < 0 \\ \{1 + ({a\over{\lambda}})e^{[\lambda 
(\sigma^- -\sigma^+)]}\}^{-1}, & ~ \mbox{if} ~\sigma^+ > 0 
\end{array}\right. ,\label{tortus}\end{eqnarray} 
where we have set $\lambda x_0 = 1$ for simplicity. 

The retarded propagator formally has the same structure than before 
but now in the tortoise coordinates. Therefore the functions
$t^\pm$ read 
\begin{equation}t^{\pm}=\partial_{\sigma^\pm}^2 \rho - 
\partial_{\sigma^\pm} \rho \partial_{\sigma^\pm} \rho,\end{equation}
and the components of the energy-momentun tensor are
\begin{equation}\langle T_{\pm\pm}\rangle = -{N\over{12}}[\partial_\pm
^2\rho - \partial_\pm \rho \partial_\pm \rho - \partial_{\sigma^\pm}^2 \rho + 
\partial_{\sigma^\pm} \rho \partial_{\sigma^\pm} \rho].\label{comptmasmas}
\end{equation}
This result is valid for all the spacetime. In particular, in the vacuum 
region $\partial_{\sigma^+}\rho$ vanishes and one obtains
\begin{equation}t^+(\sigma^+)=0.\end{equation} 
From Eq. (\ref{tortus}) we find  
\begin{equation}t^-(\sigma^-) = -{1\over{4}}\lambda^2 \left[ 1 - 
\left( 1+{a\over{\lambda}}e^{\lambda\sigma^-}\right)^{-2}\right].
\end{equation} 
which coincides with previous results \cite{cghs}.

In this section we have presented a derivation 
of the semiclassical field equations from the CTP effective action.
The covariant version of these equations contains  non-local 
terms, that become local in the conformal gauge. 

\section{Influence functional for cosmological histories}

The goal of this section will be to write the influence action for
two cosmological spacetimes and evaluate explicitly its imaginary
part. The influence functional is a matching hypersurface dependent
object. We will see that the choice of this hypersurface 
will be determinant for decoherence phenomena.

At this point, it is necessary to state precisely what does a 
``matching hypersurface" mean. We will follow closely
Ref. \cite{miguel}. Let us denote by $\cal M$ and $\tilde{\cal M}$ 
the spacetimes described
by metrics $g_{\mu\nu}^+$ and $g_{\mu\nu}^-$ respectively. We are 
assuming that both spacetimes are asymptotically flat in the past, and 
that they ``share" a spacelike hypersurface $\Sigma$. We can always  
define a hypersurface $\Sigma_{\cal M}$ in the spacetime $\cal M$ through
a relation between $t$ and $x$, say $t=k(x)$. We can also
define a hypersurface $\Sigma_{\tilde{\cal M}}$ in $\tilde{\cal M}$ by
 $\bar t={\bar k}(\bar x)$. In order to identify $\Sigma_{\cal M}$ and 
$\Sigma_{\tilde{\cal M}}$ in a common 
hypersurface $\Sigma$, we must introduce a map between
points on both hypersurfaces which follows from identifying the
local intrinsic geometry. In two dimensional dilaton gravity models,
an invariant definition of a one-geometry is provided by the
value of the dilaton field $\phi (s)$, as  a  function
of the proper distance along the hypersurface. The identification
of one-geometries therefore implies that 
for the same proper distance (measured with respect to
an arbitrary reference point) $ds^2=d\bar s^2$,
the dilaton field must have
the same value for each geometry on $ \Sigma$, i.e.,
$\phi^+(s)=\phi^-(\bar s)$. Then it follows that
$d\phi^+/ds=d\phi^-(\bar s)/d\bar s$. 

Given both spacetimes
and the function $k$ that defines the hypersurface $\Sigma_{\cal M}$ 
in  $\cal M$, the conditions imposed by the identification allow us to 
determine the function $\bar k$, and therefore the hypersurface 
$\Sigma_{\tilde{\cal M}}$ in  $\tilde {\cal M}$. If the equations 
have real solution for the function $\bar k$, then the 
hypersurface $\Sigma_{\cal M}$ ``fits'' in $\tilde{\cal M}$.

Let us now consider two cosmological metrics characterized
by the functions $\rho^+(t)$ and $\rho^-(t)$. The starting point to
compute the influence functional is the evaluation of the Green functions
$G_{ab}$. Both metrics are asymptotically flat in the past and 
conformal to Minkowski spacetime (everywhere). Therefore, the
propagators in the $in$ vacuum state have the same functional structure
than in flat spacetime. For example, the Feynman propagator is
given by

\begin{eqnarray}G_{++}(x,y)&=&i\langle 0, in\vert T{\hat f}^+(x)
{\hat f}^+(y)\vert 0,
 in\rangle\nonumber \\  
&=&{1\over{{2 \pi}^2}}\int d^2p {e^{ip(x-y)}\over{p^2 + i \epsilon}}
=- {{2\pi i}\over{{2 \pi}^2}}\int_0^\infty {dp\over{p}} e^{-i p 
(x - y)} e^{-i p\vert t_x - t_y\vert}\nonumber \\
&=& {\pi\over{2}}Sgn[\vert t_x - t_y\vert + x - y] - i Log\vert 
t_x - t_y + x - y\vert + C,\end{eqnarray}
where $C$ is an indeterminate constant (this indetermination comes
from the infrared divergence at $p\rightarrow 0$). Similar
expressions hold for the other propagators. It is 
important to note that 
in $G_{+-}(x,y)$ and $G_{-+}(x,y)$ the coordinates $x$ and $y$
correspond to different spacetimes.

\subsection{Constant time hypersurface}

Let us consider a constant time hypersurface $\Sigma_{\cal M}$ in ${\cal M}$, 
defined as $t = T$. To make the embedding we must impose 
$\phi^+=\phi^-$ on $\Sigma$. As $\phi^-$ is constant on 
$\Sigma_{\tilde{\cal M}}$, this hypersurface must also be of 
constant time ${\bar t} = {\bar T}$. After a shift of the time coordinate we
can set ${\bar T} = T$.

The CTP-effective action for cosmological metrics can be 
written as
\begin{eqnarray}S_{eff}^{CTP}&=& S_{CGHS}(\rho^+,f^+) - S_{CGHS}
(\rho^-,f^-)\nonumber \\
&-&{N\over{6 \pi}}\int_{-\infty}^T dt_x\int_{-\infty}^T dt_y 
~~\ddot{\rho}^a(t_x)~~\ddot{\rho}^b(t_y)~~\int_{-\infty}^{+\infty}
 dx\int_{-\infty}^{+\infty} dy ~~G_{ab}(x,y)
,\label{cosmoctpefa}\end{eqnarray}
where $a$ and $b$ denote the CTP branches again. We must 
compute the spatial integral of the propagator. Using dimensional 
regularization, we find 
\begin{equation}\int_{-\infty}^{+\infty} dx\int_{-\infty}^{+\infty} 
 dy ~~G_{++}= {\Omega\over{2}}~
\vert t_x - t_y\vert,\label{intprop}\end{equation}
where $\Omega$ is a global volume factor. Similar expressions hold for 
$G_{--}$ and $G_{+-}$. 

Replacing Eq. (\ref{intprop}) 
into Eq. (\ref{cosmoctpefa}), is 
possible 
to show, after some integrations by parts, that the  CTP-effective 
action for cosmological metrics is given by

\begin{eqnarray}S_{eff}^{CTP}&=& S_{CGHS}(\rho^+,f^+) - S_{CGHS}
(\rho^-,f^-)\nonumber \\
&-&{N\over{12 \pi}}\int d^2x  ~~ [\rho^+\partial_+\partial_-
\rho^+  - \rho_- \partial_+\partial_-\rho^-].\end{eqnarray}

As is immediately noted, there is not any imaginary and/or 
non-local term in this action. The only correction to the classical term
comes from the trace anomaly. The consequence of this fact 
is that the decoherence functional is identically 
one.  For the 
semiclassical approximation 
to be valid, the
decoherence functional must be diagonal for macroscopically
different spacetime geometries, even
if they coincide on a single spacelike hypersurface. Therefore,
we conclude that, due to conformal invariance, the two 
dimensional cosmological models do not have a well defined
semiclassical limit. In order to obtain such a limit, it is
necessary to break conformal invariance, as we will see in 
Section VI.

\subsection{More general hypersurfaces}

In order to show explicitly the dependence of the results with the
matching hypersurface, we will evaluate the influence functional 
for more general hypersurfaces. We will show that, even though the
action is conformally invariant, imaginary terms do appear for 
some hypersurfaces.

We must compute 

\begin{equation}\int dx\int dy \int_{-\infty}^{\Sigma} dt_x\int_{-\infty}^{
\Sigma} dt_y \ddot{\rho}^a(t_x)\ddot{\rho}^b(t_y)G_{ab}(x,y),
\label{4t}\end{equation}
where we have defined the hypersurfaces in each branch by   
\begin{equation}k(x) = T + \Delta k^+(x),\nonumber\end{equation}
and
\begin{equation}{\bar k}(x) = T + \Delta k^-(x).\nonumber\end{equation}

We will
consider that $\Delta k^+(x)$ and $\Delta k^-(x)$ are small 
fluctuations around 
the hypersurface $t=T$ and we will compute the influence functional up 
to second order in an expansion in 
powers of the fluctuations. It is obvious that the zeroth order 
gives only the trace anomaly. To 
first order the influence functional develops real terms \footnote{Note that 
these are boundary terms and therefore do not contribute to the equations 
of motion (see Eq. (\ref{supterms}))} .  The 
second order contribution is given by 

\begin{equation}\ddot{\rho}^a(T) \ddot{\rho}^b(T)\int_{-\infty}
^{+\infty} dx 
\int_{-\infty}^{+\infty} dy ~~G_{ab}(x,T;y,T)\Delta k^a(x) 
\Delta k^b(y).
\end{equation}
Introducing the corresponding propagators and performing the 
spatial integrations, we find an imaginary part proportional to

\begin{eqnarray}4 \pi i \int_0^{+\infty}{dp\over{p}}&&\left[
 \ddot{\rho}^{2+}(T) \Delta k^{+2}(p)-  \ddot{\rho}^+(T)
\ddot{\rho}^-(T) \Delta k^{+*}(p) \Delta k^-(p)\right.
\nonumber \\
&-&\left. \ddot{\rho}^-(T)
\ddot{\rho}^+(T)\Delta k^+(p) \Delta k^{-*}(p)+ 
\ddot{\rho}^{-2}(T) \Delta k^{-2}(p)\right],
\label{im}\end{eqnarray}
where $\Delta k^+(p)$ and $\Delta k^-(p)$ denote the 
Fourier transforms of the perturbation
functions. 

The basic equations describing the embedding of $\Sigma$ are

\begin{equation}\phi^+[k(x)]= \phi^+[T + \Delta k^+(x)] = \phi^-
[{\bar k}(y)]= 
\phi^-[T + \Delta k^-(y)].\label{embedding}\end{equation}
This identification
may be described by the function $y(x)$ between coordinates on 
$\Sigma$ in 
each of the spacetimes. To complete the embedding we must impose the 
intervals in each spacetime to be the same on $\Sigma$. Therefore
\begin{equation}\left[{dx\over{dy}}\right]^2={1 - \left({d{\bar k}
\over{dy}}
\right)^2\over{1 - \left({dk\over{dx}}\right)^2}}
{e^{{\rho}^-[{\bar k}(y)]}\over{e^{\rho^+[k(x)]}}}.\end{equation}

Expanding Eq. (\ref{embedding}) for small $\Delta k^+(p)$ and 
$\Delta k^-(p)$, and taking into account that $y=x+{\cal O}(\Delta k^2)$, we 
find that 

\begin{equation}\Delta k^+(x) \cong \Delta k^-(y) {\dot{\phi}^-(T)
\over{\dot{\phi}^+(T)}}\cong \Delta k^-(x) {\dot{\phi}^-(T)
\over{\dot{\phi}^+(T)}}.\label{impordess}\end{equation}
Replacing Eq. (\ref{impordess}) into (\ref{im}) the imaginary term 
can be written as

\begin{equation}4 \pi i \left[\ddot{\rho}^+(T)-{\dot{\phi}^+(T)
\over{\dot{\phi}^-(T)}}\ddot{\rho}^-(T)\right]^2\int_0^\infty 
{dp\over{p}}
\vert \Delta k^+(p)\vert^2.\end{equation}
Therefore, there will be decoherence coming 
from the small fluctuations of the hypersurface around the $t = T$ 
one, since the absolute value of the decoherence functional is given by

\begin{equation}\left\vert {\cal D}[\rho^+,\rho^-;\Sigma]\right\vert
 \approx e^
{- 4\pi\left[\ddot{\rho}^+(T)
-{\dot{\phi}^+(T)
\over{\dot{\phi}^-(T)}}\ddot{\rho}^-(T)\right]^2\int_0^\infty {dp\over{p}}
\vert \Delta k^+(p)\vert^2   }.\end{equation}

The physical interpretation of the results found in this section
is the following. The $in$ quantum state of matter is the conformal
vacuum. For $t=T$ hypersurfaces, one can choose the $out$ basis
to be the conformal vacuum on both spacetimes. Therefore, the
Bogolubov coefficients between $in$ and $out$ basis are trivial
in both geometries. The influence functional is real and there is no
decoherence. For more general hypersurfaces, one can choose as
$out$ basis the conformal vacuum in one of the spacetimes, but this
basis in general do not correspond to the conformal vacuum in the 
other. Therefore, the $in$ and $out$ basis are essentially different
in this spacetime, there is particle creation, and therefore
decoherence.

We have shown that the influence functional has an imaginary
part for some hypersurfaces, and that this imaginary
contribution vanishes for the most common hypersurfaces
of constant time. As a consequence, the absolute value of
the  decoherence functional also depends on the hypersurface. 

\section{The Dilaton loop}

In the previous sections we were using the exact effective action 
for two dimensional dilaton gravity where only the scalar matter 
fields were quantized. In order to have a more complete information 
about the quantum effects, 
we must consider the quantum fluctuations of the dilaton and the 
metric. There are some previous works where these effects have been 
taken into account, in the {\it in-out} effective action \cite{mikovic}. In 
this section we will compute the quantum correction coming from 
the dilaton field. This correction will be evaluated up to one loop 
and to lowest order in a covariant
expansion in powers of the curvature.

Starting with the classical CGHS action, we split the dilaton field in
a classical background and a small quantum fluctuation 

\begin{equation}\phi (x) =\phi_0 (x) + \hat{\phi}(x),\end{equation}   
introducing this splitting in the classical action, and dropping the 
linear terms in the fluctuation, we obtain up to quadratic order

\begin{eqnarray}S_{\phi} &=& {1\over{2 \pi}} \int d^2x \sqrt{-g(x)} 
\left\{e^{-2\phi_0}[R + 4 (\nabla \phi_0)^2 + 4 \lambda^2]\right. 
\nonumber \\
&+&\left. 4 (\nabla \psi)^2+ 2 \left[R(x) + 2 (\nabla \phi_0)^2
+ 4 \nabla^2\phi_0\right]\psi^2+8 \lambda^2 \psi^2\right\},\label{psidil}
\end{eqnarray}
where we have redefined the dilaton field as $\psi = e^{-\phi_0}\hat{\phi}$.
The action for the dilaton fluctuations $\psi$ corresponds to that 
of a massive scalar field, non conformally coupled to the curvature 
and coupled to the dilaton background. Note that this action has a
global minus sign with respect to the usual action for scalar fields. 

The Euclidean effective action can be evaluated up to quadratic  order in 
a covariant expansion in powers of the curvature \cite{rusos} as
\begin{equation}S_{eff}^\phi= {1\over{8 \pi}}\int d^2x \sqrt{g(x)}
\left[ P F_1[\Box] P-2PF_2[\Box]R+R_{\mu\nu}F_3[\Box]R^{\mu\nu}+RF_4
[\Box]R\right],\end{equation} 
where the form factors are given by

\begin{equation}F_1 = {1\over{2}}\int_0^1d\gamma ~G_\gamma^E 
,\end{equation}
\begin{equation}F_2={1\over{2}}\int_0^1d\gamma ~{(1-\gamma^2)
\over{4}}~G_\gamma^E,\end{equation}
\begin{equation}F_3={1\over{2}}\int_0^1d\gamma ~{\gamma^4\over{6}}
~G_\gamma^E,\end{equation}
\begin{equation}F_4={1\over{2}}\int_0^1d\gamma ~{(3 -6 \gamma^2- 
\gamma^4)\over{48}}~G_\gamma^E,\end{equation}
$G_\gamma^E = 
\left[{32 \lambda^2\over{1-\gamma^2}}-\Box\right]^{-1}$ 
is the Euclidean massive 
propagator, and $P$ takes into
account the nonconformal coupling $P = 2 R + 4(\nabla \phi_0)^2 + 
8 \nabla^2\phi_0$. Note that, 
as conformal invariance is broken, it is no longer possible to compute the 
effective action exactly. This effective action has a limited region of 
validity. It is applicable only
under the condition $\nabla \nabla ~{\cal R} \gg ~{\cal R}^2$, where the 
symbol ${\cal R}$ denotes any of the quantities 
$P$, $R_{\mu\nu}$ or $R$.

Using the same procedure as before, we can obtain the CTP effective 
action coming from the dilaton loop, by replacing the Euclidean massive 
propagators by the CTP ones. These propagators can be evaluated using
Riemann normal coordinates. To lowest order in the curvature, they coincide
with the usual flat propagators. Therefore it is possible to use the 
properties between the Green functions without the curved spacetime 
problems mentioned before.

The CTP effective correction from the dilaton loop is given by
\begin{eqnarray}S^{CTP}_{eff}[\phi]&=&- {1\over{2\pi}}\int d^2x\int d^2y 
~ \Delta P(x) ~ \Xi P(y)\int_0^1d\gamma ~Re~G_{++}^\gamma (x,y)~
\theta(x^0-y^0)\nonumber \\
&+&{1\over{2\pi}}\int d^2x\int d^2y 
~ \Delta P(x)~ \Xi R(y)\int_0^1d\gamma ~{(1 - \gamma^2)\over{2}} 
~Re~G_{++}^\gamma (x,y)~\theta(x^0-y^0)\nonumber \\
&-&{1\over{2\pi}}\int d^2x\int d^2y ~ \Delta R(x)
~ \Xi R(y)\int_0^1d\gamma ~{(1 - 2\gamma^2+3\gamma^4)\over{16}} 
~Re~G_{++}^\gamma (x,y)~\theta(x^0-y^0)\nonumber\\
&+&{i\over{4\pi}}\int d^2x\int d^2y 
~ \Delta P(x)~ \Delta P(y)\int_0^1d\gamma ~Im~G_{++}^\gamma (x,y)
\nonumber \\
&-&{i\over{4\pi}}\int d^2x\int d^2y 
~ \Delta P(x)~ \Delta R(y)\int_0^1d\gamma ~{(1 - \gamma^2)\over{2}}
~Im~G_{++}^\gamma (x,y)\nonumber \\
&+&{i\over{4\pi}}\int d^2x\int d^2y ~\Delta R(x)
~ \Delta R(y)\int_0^1d\gamma ~{(1 - 2\gamma^2+3\gamma^4)\over{16}}
Im~ G_{++}^\gamma 
(x,y),\label{dilatonloop}\end{eqnarray}
where with $\Delta$ and $\Xi$ we are denoting a half of the difference 
and sum, respectively, between the fields in each of the CTP branches; 
 and where
\begin{equation}G_{++}^\gamma (x,y)=\int {d^2p\over{(2\pi)^2}}{e^{ip(x-y)}
\over{p^2+{32 \lambda^2\over{1-\gamma^2}}-i\epsilon}}.\end{equation}
Owing to the global minus sign in the dilaton action, the CTP effective 
action for the dilaton has a global minus sign in its real part.

Finally, the total effective action including the dilaton loop can be 
written as Eq. (\ref{explctp}) (that is an exact result) plus 
Eq. (\ref{dilatonloop}) (valid up to second order in powers of 
the curvature). Then, the absolute value of the decoherence functional
has two contributions: the imaginary part of boundary terms in 
Eq. (\ref{supterms}) plus the imaginary part of the dilaton 
effective action

\begin{equation}\left\vert {\cal D}[\rho^+,\rho^-]\right\vert 
\approx e^{-\mbox{Im}~\left[\mbox{boundary terms}~ + ~S_{eff}^{\phi}\right]}
.\label{compltDF}
\end{equation}
In particular, in the cosmological case and when the common hypersurface is 
defined by $t=T$, the boundary terms vanish, and Eq. (\ref{compltDF}) shows
that decoherence still appears if one includes the dilaton loop in the 
evaluation of the effective action.

It is clear that a complete model should include the graviton loop too. Here 
we considered only the contribution of the dilaton. In the more realistic 
case where the quantum fluctuacions coming from the metric are 
also included, an extra imaginary part will be obviously present, but 
our conclusion about the appearance of decoherence effects will 
be unaffected. It is also interesting to note that one can obtain 
the covariant semiclassical field equations taking the variation of 
action (\ref{dilatonloop}) with respect to the metric.  
 
Conformal invariance can, of course, be broken in many other ways, for
example by considering massive and/or non-minimally coupled matter fields. The
classical action for such field is 

\begin{equation}S_f = -{1\over{2}}\int ~d^2x ~\sqrt{g} ~[(\nabla f)^2 +
(m^2 + \xi R)f^2],\end{equation}
which is similar to the action for the quantum fluctuations of 
the dilaton (Eq. (\ref{psidil})).
The CTP effective action can be computed again using a covariant 
expansion in powers of the curvature, and one obtains a result 
similar to Eq. (\ref{dilatonloop}).

\section{Final Remarks}

In this paper we have computed the influence functional for two dimensional
models of dilaton gravity. When only conformal matter fields are quantized,
 the influence functional can be computed exactly. This is, of course, a 
consequence of the conformal invariance. We have also shown that the 
influence functional depends strongly on the matching hypersurface.
In particular, in the conformal gauge the influence action can be written
as the difference between the Liouville actions for the metrics on the +/$-$ 
branches plus an integral over $\Sigma$.

We used the influence action to derive the covariant form of the 
semiclassical field equations. These equations are real, causal and non-local, 
and become local in the conformal gauge. The derivation is non trivial due 
to the dependence of the propagators with the metric. It is not only
of academic interest, since the procedure can be generalized to cases 
when conformal invariance is broken, i.e. when it is difficult to
 evaluate the $\langle T_{\mu\nu}\rangle $ using conservation laws and the
 trace anomaly.

We have studied the quantum to classical transition in cosmological 
backgrounds. We have shown that the influence functional does not 
contain imaginary parts for some matching hypersurfaces. Therefore the
semiclassical approximation is not valid in the conformal case. 

We have also pointed out that the semiclassical approximation can be 
recovered by including the dilaton loop (and eventually quantum 
fluctuations of the metric), since in this case conformal invariance is 
broken. Two remarks are in order. The first one is that the quantum 
fluctuations of the dilaton also imply the validity of the semiclassical 
approximation for the dilaton background. This is important, since the 
``geometry'' in the two dimensional models is determined by the metric and
the dilaton. Moreover, when the two dimensional models are obtained by 
restricting the four dimensional Einstein-Hilbert action to metrics
with spherical symmetry, $\phi$ is part of the geometry since 
$e^{-2\phi}$ is 
the radius of the 2-sphere. The
second remark is that the dilaton and the metric loops are usually neglected
 by invoking the large N limit. However, we have seen that, no matter how 
large N is, there is no decoherence unless conformal invariance is 
broken. The dilaton and graviton loops are crucial in this sense. 

It is worth to note that the existence of an imaginary part in the influence
action is a necessary but not a sufficient condition for the validity of the
 semiclassical limit. In order to
show semiclassicality we must have sufficient decoherence while maintaining 
classical correlations. This requires a more detailed analysis that is 
beyond the scope of this paper.
 
In Section VI we have noted that the action for the dilaton fluctuations 
has a global minus sign with respect to the action for the matter 
fields. It would be interesting to investigate the effect of these 
fluctuations in the semiclassical field equations.  

Finally, we would like to comment on related works about the validity of 
the semiclassical approximation in these models. In Refs. \cite{miguel} and 
\cite{boseparkerpeleg}, this problem has been investigated in the vicinity
 of the black hole horizon. The main idea in those papers was the following. 
If the semiclassical approximation were a good one, the wave functional
 of the quantum fields should not depend very strongly on the black hole 
mass. Indeed, if we consider two different spacetimes, one describing the 
collapse of a black hole with mass $M$, the other with mass $M + \Delta M$, 
similar wave functionals at early times should not be ``too different'' after
the black hole is formed (if $\Delta M$ is sufficiently small). In order
to compare both wave functionals, one can embedd a spacelike hypersurface
$\Sigma$ in both spacetimes, and compute their inner product on $\Sigma$. It 
has been shown that, for certain hypersurfaces, this inner
 product is arbitrary small for the classical collapsing geometries 
\cite{miguel}, while it is of order one if quantum backreaction is 
included in the collapse \cite{boseparkerpeleg}. 

The alert reader should have noticed that the inner product defined in
Refs. \cite{miguel} and \cite{boseparkerpeleg} is exactly the 
influence functional $e^{i S_{eff}^{CTP}}$, 
evaluated at the geometries of collapsing black hole of masses $M$ and 
$M+\Delta M$, with matching hypersurface $\Sigma$. The main technical 
difference between their calculation and ours is that they have worked in 
the 
Schr\"odinger picture, while we worked in the Heisenberg picture. 
In principle, all the results about the validity of the semiclassical 
approximation near the horizon of a black hole should be contained in 
the boundary terms that appear in the influence functional (Eq. 
(\ref{supterms})). 
Indeed, the result (\ref{supterms}) is completely general, valid for any 
pair of metrics and any matching hypersurface. However, as we already 
pointed out, a complete analysis of this problem should include
 dilaton and metric fluctuations, since the semiclassical approximation 
in two dimensional models will be valid if both the metric and the dilaton
decohere.

\acknowledgments
We would like to thank Bei Lok Hu and Miguel Ortiz for many useful 
discussions. This research was supported by Universidad de Buenos Aires, 
Consejo Nacional de Investigaciones Cient\'\i ficas y T\'ecnicas (CONICET), 
and by Fundaci\'on Antorchas.

\end{document}